\documentclass[fleqn,10pt]{wlscirep}
\usepackage[utf8]{inputenc}
\usepackage[T1]{fontenc}

\usepackage{lipsum}

\usepackage{xcolor}
\usepackage{epsfig,amsmath}
\usepackage{graphicx}
\usepackage{dcolumn}
\usepackage{bm}
\usepackage[]{amsmath}
\usepackage{braket}
\usepackage{amsfonts,amsmath}
\usepackage{epsfig,amsmath}
\usepackage{graphicx}
\usepackage{dcolumn}
\usepackage{bm}
\usepackage{bbold}
\newcommand{\beq}{\begin{equation}}
\newcommand{\eeq}{\end{equation}}
\newcommand{\beqa}{\begin{eqnarray}}
\newcommand{\eeqa}{\end{eqnarray}}

\newcommand{\la}{\langle} 
\newcommand{\ra}{\rangle}

\def\oe#1{{ Opt.\ Express} {\bf#1}}

\title{Breaking the Diffraction Barrier for Passive Sources: Parameter-Decoupled Superresolution Assisted by Physics-Informed Machine Learning}

\author[1]{Abdelali Sajia}
\author[1,2]{Bilal Benzimoun}
\author[1]{Pawan Khatiwada}
\author[1]{Guogan Zhao}
\author[1,*]{Xiao-Feng Qian}
\affil[1]{Center for Quantum Science and Engineering, and Department of Physics, Stevens Institute of Technology, Hoboken, New Jersey 07030, USA}
\affil[2]{Department of Physics, Clark University, Worcester, Massachusetts 01610, USA}

\affil[*]{corresponding author: xqian6@stevens.edu }

\begin{abstract}
We present a parameter-decoupled superresolution framework for estimating sub-wavelength separations of passive two-point sources without requiring prior knowledge or control of the source. Our theoretical foundation circumvents the need to estimate multiple challenging parameters such as partial coherence, brightness imbalance, random relative phase, and photon statistics. A physics-informed machine learning (ML) model (trained with a standard desktop workstation), synergistically integrating this theory, further addresses practical imperfections including background noise, photon loss, and centroid/orientation misalignment. The integrated parameter-decoupling superresolution method achieves resolution 14 and more times below the diffraction limit (corresponding to $\sim 13.5$ nm in optical microscopy) on experimentally generated realistic images
with $>82\%$ fidelity, performance rivaling state-of-the-art techniques for actively controllable sources. Critically, our method’s robustness against source parameter variability and source-independent noises enables potential applications in realistic scenarios where source control is infeasible, such as astrophysical imaging, live-cell microscopy, and quantum metrology. This work bridges a critical gap between theoretical superresolution limits and practical implementations for passive systems.
\end{abstract}
\begin{document}

\flushbottom
\maketitle
 
\thispagestyle{empty}

\section{Introduction}

For over a century, the Abbe resolution limit has been a fundamental barrier, constraining the smallest resolvable separation between two points in various imaging, sensing, and precision measurement applications \cite{abbe1873beitrage, born1999principles}. This limit, usually expressed as $d\approx\lambda/2{\rm NA}$, corresponds to approximately 180 nm under typical microscope numerical aperture ${\rm NA}\approx 1.4$ for visible (e.g., green) light wavelength 
$\lambda \approx 500$ nm. According to Rayleigh's criterion, the limit arises because, as two sources come closer together, their overlapping blurred signal at the imaging plane become increasingly difficult to distinguish through direct intensity measurements \cite{rayleigh1879xxxi, rayleigh1896xv, goodman2005introduction, kolobov2000quantum}. Consequently, this minimum resolvable distance is fundamentally tied to the width 
$\sigma$, which is twice of the full width at half maximum (FWHM) $\approx 0.255 \lambda/{\rm NA}$, of the point spread function (PSF) for the point source. Then the PSF width also quantifies the system’s diffraction-limited resolution such that $d \approx \sigma$ \cite{goodman2005introduction,tsang2016quantum,scheiderer2025minflux}.

Remarkably, various superresolution microscopy (SRM) techniques for active sources, those that can be partially controlled or pre-engineered, have been developed to overcome the optical resolution limit. Methods such as stimulated emission depletion microscopy (STED) \cite{hell1994breaking}, structured illumination microscopy (SIM) \cite{gustafsson2000surpassing}, photo-activated localization microscopy (PALM) \cite{betzig2006imaging}, stochastic optical reconstruction microscopy (STORM) \cite{rust2006sub}, and Minimal (fluorescence) photon flux microscopy (MINFLUX) \cite{scheiderer2025minflux}, have achieved significant improvements, enhancing resolution from approximately 200 nm to as fine as 10 nm \cite{galbraith2011super, prakash2022super, chen2023superresolution, zhao2022sparse, fang2023fluorescent}. However, these methods are ineffective in more general scenarios involving passive sources, those that cannot be controlled or must remain undisturbed \cite{stelzer2002beyond, bettens1999model, van2002high, ram2006beyond, chao2016fisher, kolobov2000quantum, stelzer2000uncertainty}. Examples include resolving two distant stars, distinguishing two living biological units that would perish if interfered with, or differentiating two unknown sources.


Recently, a novel approach based on quantum spatial mode demultiplexing has demonstrated the potential to achieve sub-wavelength superresolution for passive point sources without requiring control or manipulation of the source \cite{tsang2015quantum, tsang2016quantum, tsang2019resurgence, hradil2019quantum, hradil2021exploring, larson2018resurgence, wadood2021experimental, liang2021coherence, liang2023quantum,santamaria2023OE,paur2016O, darji2024robust,hervas2024optimizing, thachil2023achieving, tsang2019resolving, sorelli2024multimode, sajia2023superresolution}. However, this coefficient-based state decomposition method is limited to the idealized scenario of two equally bright and completely incoherent sources \cite{tsang2016quantum, hradil2019quantum, larson2018resurgence, schlichtholz2024superresolution, schlichtholz2025superresolving, napoli2019towards}. Efforts to address the challenge of unbalanced brightness have been proposed \cite{sajia2022superresolution, linowski2023NJP, Rehacek2017PRA, li2024quantum, zhang2024superresolution, santamaria2024single}.

On another front, machine learning (ML) techniques have been employed to reconstruct super-resolved images using spatial mode decomposition \cite{kudyshev2023machine, frank2023O, Rouviere2024O}. Nevertheless, these approaches remain constrained to incoherent images. Particularly, achieving higher resolution requires using a greater number of spatial modes, as well as detailed amplitude and phase information for each mode \cite{an2020AO}. These requirements significantly increase the demands on measurement precision and control, resulting the method more susceptible to imperfections.

To date, comprehensive treatments addressing major source-dependent practical challenges, i.e., partial coherence, unbalanced brightness (or balanceness), random phase, unknown photon statistics, remain absent. Additionally, several inevitable source-independent factors, including background noise, detection inefficiency (or photon loss), source center-point/orientation detection misalignment, have yet to be thoroughly investigated.

\begin{figure*}[h!]
\centering
\includegraphics[width=17cm]{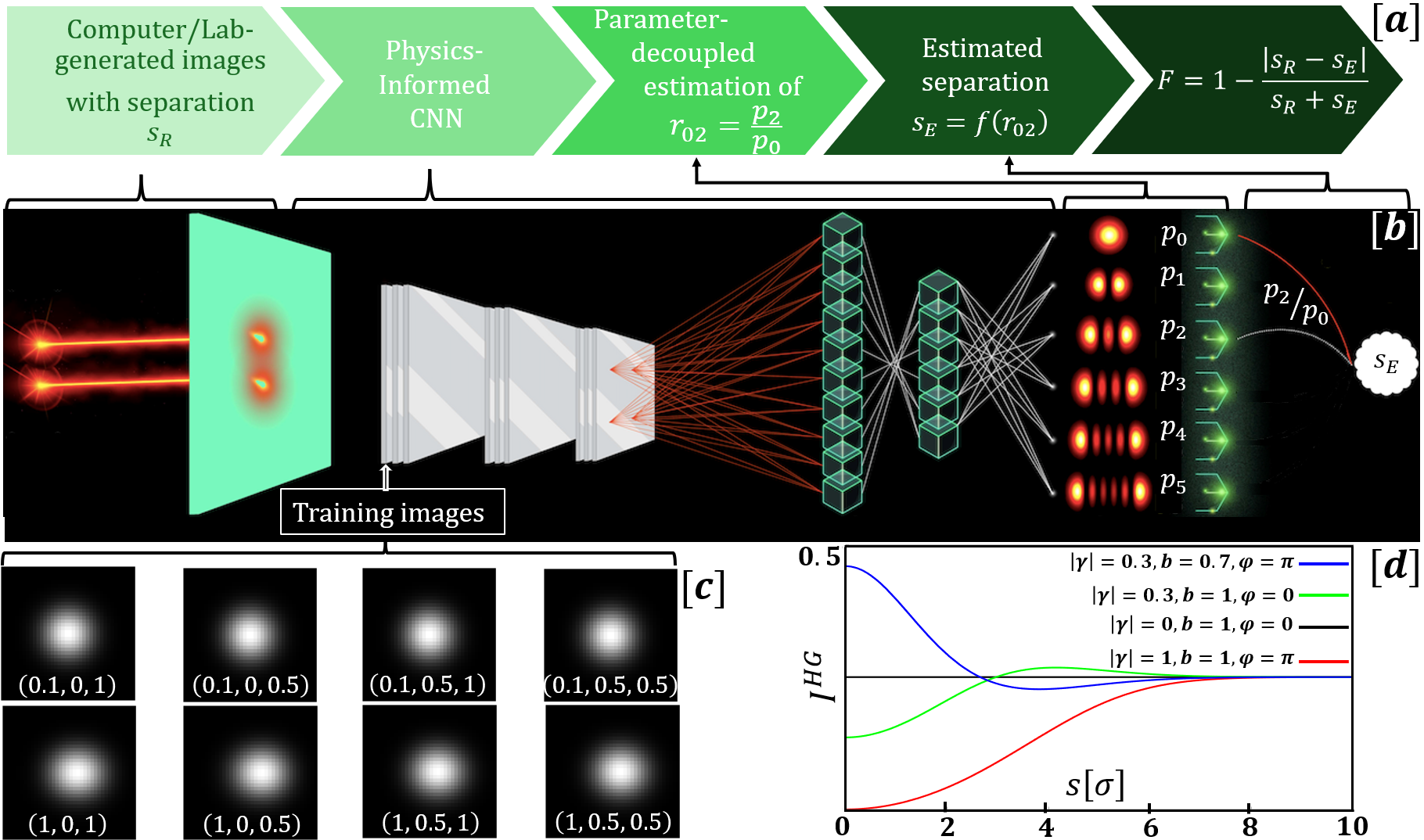}
\caption{Parameter-decoupled superresolution framework for passive sources. (a) Conceptual workflow of the parameter-decoupled protocol for estimating and evaluating two passive point-source separation. (b) Architecture of the physics-informed CNN, trained to infer the separation $s$ from input images. (c) Example training images with varying sets of parameters ($s$, $|\gamma|$, $b$) with the separation $s$ in the unit of $\sigma$. (d) Fisher information $I^{HG}$ for estimating $s$. Unlike conventional methods, $I^{HG}$ remains finite at $s=0$ (except for the idealized case of perfect destructive interference, red curve), ensuring robust superresolution across realistic experimental conditions.}
\label{roadmap}
\end{figure*}

In this article, we theoretically develop and experimentally demonstrate a systematic parameter-decoupling approach to superresolve generic passive two-point sources. Our method bypasses the need to estimate practical parameters including partial coherence, brightness imbalance, random relative phase, and photon statistics by leveraging the probability/intensity ratio of contributing spatial modes. This eliminates the requirement for detailed phase, amplitude, and photon number information of each individual mode, thereby significantly simplifying measurement complexity. We further integrate this theoretical framework with a Convolutional Neural Network (CNN) trained specifically on the developed model. The ML approach effectively addresses additional practical challenges, including background luminosity noise, detection inefficiency (or photon loss), and center-point/orientation misalignment. Finally, we validate our model with experimentally generated images from two different mechanisms of realistic two-point sources, achieving over 82\% fidelity for distances more than 14 times below the conventional resolution limit $\sigma$ (or $\approx d$). These results represent a significant advancement in superresolution for practical passive sources, with potential applications in astrophysical and biological imaging.


We model arbitrary two-point passive sources with a generic statistical mixture of two spatial states of photons (or classical light fields), represented by a non-normalized density matrix  $\rho$ (or a statistical field), see the Methods section for details. The system’s behavior depends on several unknown two-source parameters: the separation $s$, partial coherence $|\gamma|$, probability (or intensity) imbalance $b$, random relative phase $\varphi$, and photon statistics. Conventional measurements, which rely on the pixel by pixel probability (or intensity) $Tr [\rho]$, and specialized spatial mode decomposition techniques, such as projections onto the $q$-th Hermite-Gaussian (HG) modes with probability $p_q$ \cite{tsang2016quantum}, inherently couple all four parameters ($s,|\gamma|,b,\varphi$) and are further influenced by photon statistics. This unwanted coupling makes it practically infeasible to extract reliable estimates of the separation $s$ from these measurements alone.



To overcome the multi-parameter estimation challenge encountered in prior studies, we propose a parameter-decoupling protocol based on the ratio of probabilities (or intensities) between even or odd HG spatial modes. Specifically, we employ $r_{mn}=p_n/p_m$, where $m,n$ are even (or odd) indices (e.g., 0,2,4,...) and $p_n$ ($p_m$) are corresponding projection probabilities of the two-source signal on the $n$-th HG mode. Crucially, this ratio depends exclusively on two-point separation $s$, i.e., 
\beq
r_{mn}=\frac{p_n(|\gamma|,b,\varphi,s)}{p_m(|\gamma|,b,\varphi,s)}\equiv r_{mn} (s),
\eeq
and it is also independent of total photon numbers. This approach decouples $s$ from other parameters, ensuring that the Fisher information remains practically non-zero even as $s \rightarrow 0$, thereby overcoming the limitations of traditional multi-parameter estimation. Detailed analyses of this ratio $r_{mn}$ and Fisher information are provided in the Methods section.

Figure \ref{roadmap}(a) illustrates the framework of our parameter-decoupled superresolution method for passive sources. Target images are fed into a physics-informed CNN [schematically shown in Fig. \ref{roadmap}(b)], which is trained using the parameter-decoupled ratio $r_{20}$ to estimate the separation $s_E$. Performance is evaluated via a fidelity measure $F$, quantifying the agreement between $s_E$ and the true separation $s_R$. Representative training images with varying parameter combinations of ($s$, $|\gamma|$, $b$) are shown in Fig. \ref{roadmap} (c). Figure \ref{roadmap} (d) demonstrates the Fisher information $I^{HG}$ associated with the ratio $r$ for the separation $s$. Notably, $I^{HG}$ remains finite at $s=0$ except in the idealized case of perfect destructive interference, confirming the robustness of our protocol.

\section{Results}


Through physics-informed training, the CNN model achieves 99\% accuracy in resolving the probability ratio of Hermite-Gaussian (HG) modes from testing samples, demonstrating high quality classification capability to unseen data. Detailed CNN training procedures and treatments are described in the Methods section. However, this classification accuracy does not reflect directly the precision of super-resolution separation. To rigorously evaluate the superresolution reconstruction fidelity, we introduce a normalized metric quantifying the agreement between the estimated separation $s_E$ and the ground-truth separation $s_R$ in testing images, i.e., 
\begin{align}
\label{fidelity}
 F=1-\frac{|s_{E}-s_{R}|}{s_{E}+s_{R}}.
\end{align}
Obviously, $F\in[0,1]$, serving as a symmetric measure of relative error, achieving $F=1$ for perfect agreement ($s_E=s_R$) and decreasing proportionally to the discrepancy between estimates and ground truth. In the limiting case, $F\rightarrow0$ due to either $s_E\ll s_R$ or $s_E\gg s_R$, indicating the worst measurement estimation with a maximum possible error. 

To rigorously evaluate the fidelity of the physics-trained CNN model, we validate its performance across two distinct categories of test data: synthetic (computer-generated) images and experimental (lab-acquired) images. This dual-validation approach ensures robustness in both idealized and real-world scenarios.

\subsection{Computer-generated data tests}
Computer-generated testing images are used to explore the robustness of our CNN model in three aspects: (a) source-dependent imperfections, (b) source-independent factors, and (c) training image properties.

\subsubsection{Robustness against unknown source-dependent parameters}
Here we demonstrate superresolution fidelity behaviors of the CNN model for computer-generated data under multiple configurations of four practical source-dependent parameters: separation $s$, partial coherence $|\gamma|$, balanceness $b$, and relative phase $\varphi$.

\begin{figure}[h!]
\centering
\includegraphics[width=15.5cm]{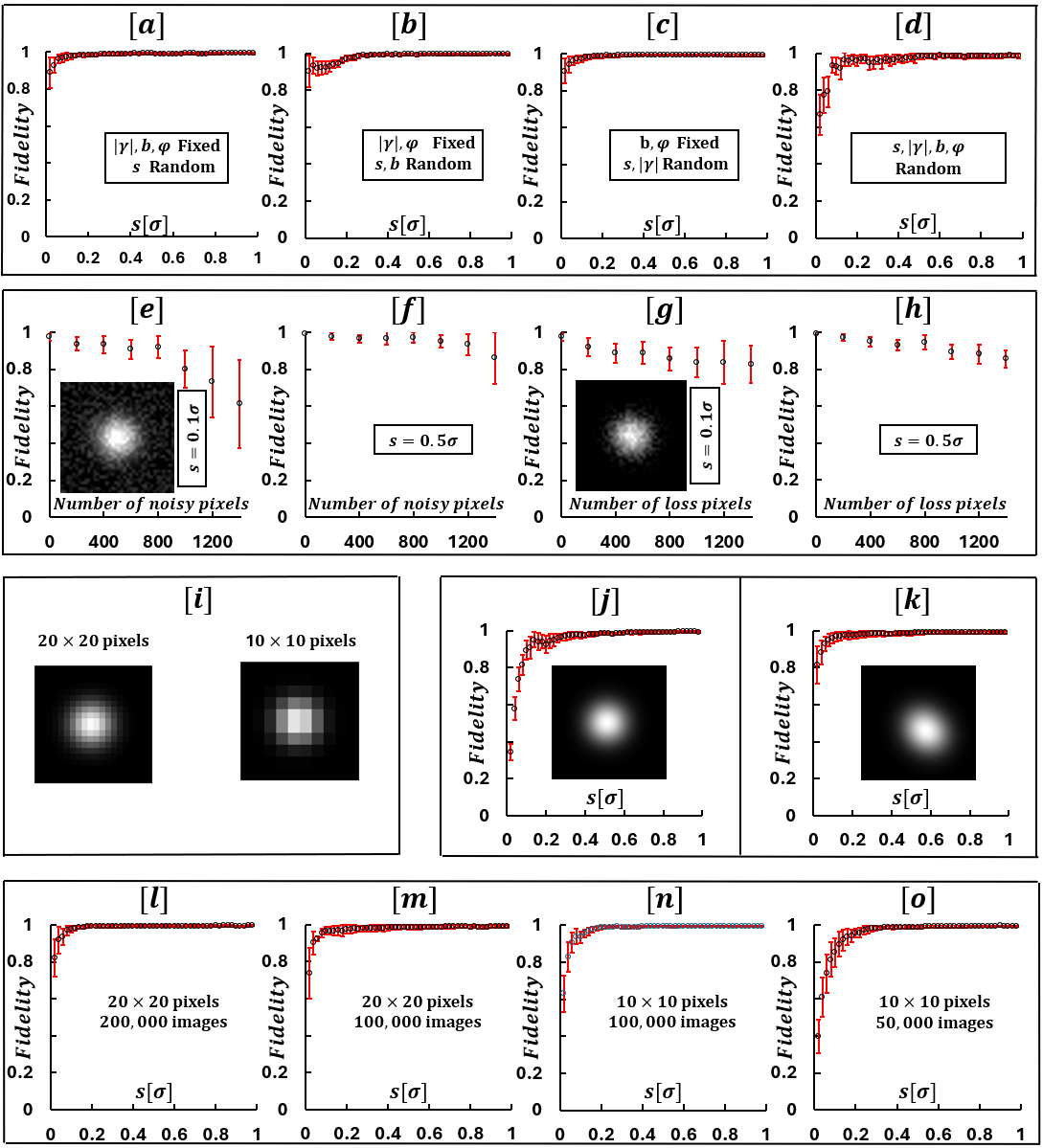}
\caption{Separation estimation fidelity of the physics-informed CNN model for computer-generated data. Effects of source-dependent parameters: (a) illustrates the case when only $s$ is unknown while other parameters are fixed $\gamma=0$, $b=1$, $\varphi=0$, (b) represents when $s$ and $b$ are unknown, while $\gamma=0$, $\varphi=0$ are fixed, (c) shows the case when both $s$ and $\gamma$ are unknown, while $b=1$ and $\varphi=0$ are fixed, and (d) depicts when all four parameters are unknown and randomized. Effects of source-independent factors: (e) and (f) illustrate the fidelity with respect to the number of polluted pixels with luminosity noise for separations $s=0.1\sigma$ and $0.5 \sigma$ respectively. A sample image with 800 noisy pixels is shown for separation $s=0.1\sigma$. Panels (g) and (h) illustrate the fidelity with respect to the number of polluted pixels with photon loss for separations $s=0.1\sigma$ and $0.5 \sigma$ respectively. A sample image with 1000 affected pixels is shown for separation $s=0.1\sigma$. Panel (j) illustrates the fidelity with respect to $s$ for data images with 5\% centroid misalignment and a sample data image, and panel (k) illustrates fidelity with respect to $s$ for data images with arbitrary orientation misalignment and a sample data image. Effects of training properties: Panel (i) illustrates two data images for separation $s=0.1\sigma$ with low resolutions of ($10\times10$) and ($20\times20$) pixels, respectively. Panels (l-n) illustrate fidelity behaviors for different number of training images and different number of pixels per image.}
\label{multiple-parameter noise}
\end{figure}

Fig.~\ref{multiple-parameter noise} (a) illustrates the fidelity behavior for an ideal scenario where three parameters, $\gamma = 0$, $b= 1$, and $\varphi = 0$, are fixed, while the separation $s$ is fully randomized. In Fig.~\ref{multiple-parameter noise} (b), both $s$ and $b$ are randomized and treated as unknown variables, with 
$\gamma = 0$, $\varphi = 0$ remaining fixed. Fig.~\ref{multiple-parameter noise} (c) extends this analysis to a case where $s$ and $|\gamma|$ are randomized and unknown during training and testing, while $b = 1$ and $\varphi = 0$ are held constant. Notably, across all three configurations, the fidelity $F$ rapidly converges to nearly unity as the separation $s$ increases. Even for sub-wavelength separations as small as
$s=0.02\sigma$ (50 times below the conventional resolution limit, $\sim$3.6 nm in typical optical microscopy setups), the system achieves an average fidelity of $F=0.9$, demonstrating robustness against parameter uncertainties.

Fig.~\ref{multiple-parameter noise} (d) depicts the more demanding scenario, where all source-dependent parameters, i.e., separation $s$, coherence $|\gamma|$, balanceness $b$, and phase $\varphi$, are randomized and unknown. Despite this full parameter indeterminacy, the superresolution fidelity retains values above $F>0.82$ for all separations $s\ge 0.06\sigma$ ($\sim$10.8 nm in typical optical microscopy setups), demonstrating resilience even under extreme uncertainty.

These results demonstrate the strong reliability of the physics-informed CNN model in estimating two-source separations. In particular, the relatively small error bar highlights the stability of our model.

\subsubsection{Robustness against source-independent imperfections}
Here we demonstrate the influence of source-independent factors on the fidelity of our CNN model's separation estimation for simulated computer-generated images. By fixing all source-dependent parameters $|\gamma|=0$, $b=1$, $\varphi=0$, we isolate and highlight the effects arising purely from source-independent variables.

All training and testing samples are scaled such that pixel intensities lie within the range $[0,255]$. Background noise is introduced by adding a random value between 10 and 50 to a set of randomly selected pixel’s intensity before rescaling. This corresponds to 4\%–20\% of the maximum signal intensity (relative to the 255-scale), simulating typical real-world noise levels encountered in common scenarios. We begin by randomly selecting 200 pixels for artificial pollution augmentation. In each subsequent iteration, we incrementally expand the affected area by adding 200 additional pixels, progressively increasing the spatial coverage of the pollution simulation.

Fig.~\ref{multiple-parameter noise} (e) and (f) illustrate the fidelity trends of our CNN model under increasing numbers of noisy pixels at fixed separations of $s=\sigma/10$ and $\sigma/2$, respectively. The model demonstrates remarkable robustness: at 
$s=\sigma/10$, fidelity remains as high as $F=0.9$ even with 800 noisy pixels (over 55\% of the total 
$38\times38=1444$ pixels). A sample noisy data image is illustrated in Fig.~\ref{multiple-parameter noise} (e) with 800 polluted pixels. At $s=\sigma/2$, this robustness extends further to 1200 noisy pixels (83\% of all pixels). A sample noisy data image is illustrated in Fig.~\ref{multiple-parameter noise} (g) with 1000 polluted pixels. Apparently, larger separation is more robust against noise.

In a complementary test, we investigate photon loss effects by simulating pixel-wise luminosity reduction. For randomly selected pixels, we subtract a random value between 10 and 50 from their intensity values, ensuring a minimum luminosity threshold of zero. Fig.~\ref{multiple-parameter noise} (g) and (h) depict the fidelity trends under varying numbers of photon-depleted pixels at fixed separations $\sigma/10$ and $\sigma/2$, respectively. Our model exhibits strong resilience to photon loss, with fidelity decreasing only marginally as the number of affected pixels increases. At $s=\sigma/10$, fidelity remains high ($F>0.9$) for 600 photon-depleted pixels (41\% of the total 1444 pixels), and at $s=\sigma/2$, this robustness extends to 1000 pixels (69\% of all pixels). Notably, larger separations demonstrate greater robustness against photon loss (or detection inefficiency) as well.

We further demonstrate the impact of (detection-caused) two-source centroid misalignment and orientation misalignment on the fidelity of separation estimation. While the CNN model inherently achieves highly accurate image recognition, automatically localizing centroids and orientations with sub-pixel precision even at small separations $s=0.02\sigma$, equivalent to 3.6 nm in typical optical microscopy, we rigorously test its robustness by artificially introducing errors. 

For centroid misalignment, we apply a forced 5\% shift (1 pixel) to testing images, see a sample image in Fig.~\ref{multiple-parameter noise} (j). Fig.~\ref{multiple-parameter noise} (j) also shows that the CNN model maintains an average fidelity $F>0.8$ for separations $s>0.08\sigma$ (equivalent of 14.4 nm), despite being trained with a 7.5\% random centroid shift to enhance tolerance.

To evaluate orientation misalignment, we generate testing images (Fig.~\ref{multiple-parameter noise} (k) illustrates a sample image) with random angular deviations up to $\pi/2$ from the true orientation. Fig.~\ref{multiple-parameter noise} (k) reveals that this orientation error has negligible impact on fidelity. Compared to the orientation error-free case in Fig.~\ref{multiple-parameter noise} (a), performance shows only a minor reduction, underscoring the model’s robustness to orientation misalignment.

The combination of these source-independent factors, background luminosity noise, photon loss, centroid and orientation misalignment, with source-dependent practical parameters will be tested by real experimental data in the following.

\subsubsection{Effects of training image properties}

Beyond source-dependent and source-independent imperfections, the fidelity of our CNN-based separation estimation model is also influenced by properties of the training dataset itself. We investigate two critical factors: the number of training images and the image resolution (or number of pixels per image). To isolate effects purely attributable to training properties, we fix all non-separation source-dependent parameters ($|\gamma|=0$, $b=1$, $\varphi=0$) and exclude source-independent imperfections during this analysis.

Fig.~\ref{multiple-parameter noise} (l)-(o) illustrate fidelity behaviors with respect to $s$ for four different configurations of training properties. Panel (l) shows average fidelity surpasses 0.9 after $s> 0.04\sigma$ for training with 200,000 images ($20\times 20$ pixels). Panels (m) and (n) correspond to 100,000-image training with $20\times 20$ and 
$10\times 10$ pixels, respectively. Both cases achieve fidelity above 0.9 for $s> 0.06\sigma$. Panel (o) presents the case of 50,000 training images $10\times 10$ pixels, where fidelity exceeds 0.9 for $s> 0.12\sigma$. Sample testing two-point source images for $20\times 20$ and $10\times 10$ pixels are shown in Fig.~\ref{multiple-parameter noise} (i).

These results highlight two key trends. First, for a fixed pixel resolution, a larger number of training images improves fidelity, as evident in the comparisons between panels (l) and (m), as well as between panels (n) and (o). Second, when the number of training images per pixel remains constant, higher pixel resolution enhances fidelity, as shown by panels (l) and (o). Therefore, increasing pixel resolution does not always guarantee better performance when the total number of training images is fixed, as shown in panels (m) and (n). In this scenario, higher resolution leads to fewer training images per pixel, creating a trade-off between resolution and sample density that results in comparable performance for both cases.

\subsection{Lab-generated data tests}

\begin{figure}[h!]
\centering
\includegraphics[width=16cm]{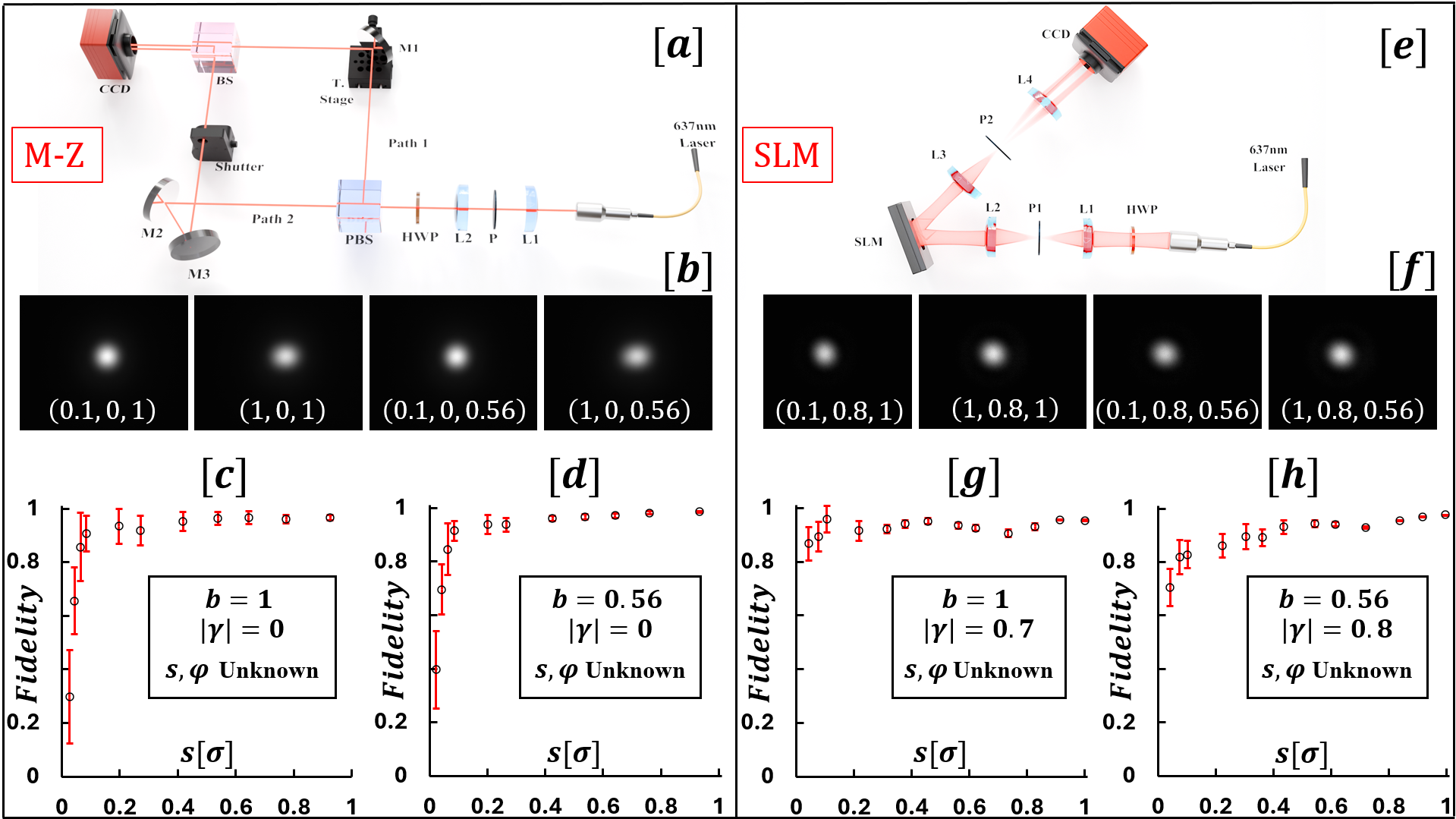}
\caption{Separation estimation fidelity of the physics-informed CNN model for experimental data. MZ Interferometer Experiments: (a) schematic illustration of the MZ interferometer setup to generate two-point source data images; (b) experimental raw data images for varying sets of parameters ($s$, $|\gamma|=0$, $b$), with fixed phase $\varphi=0$; (c) and (d) illustrate fidelity behaviors for incoherent ($|\gamma|=0$) data under balanced ($b=1$) and partially balanced ($b=0.56$) conditions, respectively. SLM Setup Experiments: (e) schematic illustration of the SLM setup to generate two-point source data images; (f) experimental raw data images for varying sets of parameters ($s$, $|\gamma|=0$, $b$) with fixed phase $\varphi=0$; (g) shows the fidelity behaviors for balanced ($b=1$) and partially coherent ($|\gamma|=0.7$) SLM testing data; (h) illustrates the fidelity behaviors for partially balanced ($b=0.56$) and partially coherent ($|\gamma|=0.8$) SLM testing data.}
\label{experiment}
\end{figure}

We experimentally validate our CNN model using two distinct methods to generate realistic two-point source, i.e., utilizing a Mach-Zehnder (MZ) interferometer and a spatial light modulator (SLM), see schematic illustrations in Fig.~\ref{experiment} (a) and (e) respectively. The resulting images inherently incorporate all practical imperfections: source-independent factors (e.g., background noise, photon loss, centroid/orientation misalignment, etc.) and source-dependent parameters (partial coherence $|\gamma|\neq 0$, partial balanceness $b\neq 1$, random separation $s$, and random relative phase $\varphi$). Fig.~\ref{experiment} (b) and (f) display representative two-point source images (raw, pre-centering/resizing) captured from the MZ and SLM setups respectively.

Prior to CNN analysis, raw camera-captured images undergo two pre-processing steps. (1) Separation calibration: the actual separation between sources is determined by localizing each source individually. For the MZ setup, this is achieved via a mechanical shutter and precision translation stage; for the SLM, separation is controlled through computer-generated holograms.
(2) Standardization: images are cropped to center the two-source signal and resized to $38\times38$  pixels, matching the format of the training dataset. This step may introduce some centroid and orientation misalignment. For example, images from the SLM setups (see Fig.~\ref{experiment} (f)) are relatively very tiny compared to the entire camera screen, and the entire signal is off from the center of the camera. It is remarkable to note that even with such offsets, our physics-trained CNN model is still performing at a very high quality level, as shown below.

Fig.~\ref{experiment} (c) and (d) show the fidelity trends for incoherent ($|\gamma|=0$) testing data from the MZ setup under balanced ($b=1$) and partially balanced ($b=0.56$) conditions, respectively. Both cases exhibit similar fidelity behavior, surpassing $F>0.85$ for separations $s\gtrsim 0.064\sigma$ (11.3 nm in typical optical microscopy). Fig.~\ref{experiment} (g) and (h) depict results for SLM-generated data. For partially coherent but balanced sources ($|\gamma|=0.7$, $b=1$) in panel (g), the CNN model achieves $F>0.87$ at $s>0.04\sigma$ (7.2 nm in typical optical microscopy). Remarkably, this robustness persists even under the simultaneous partial coherence and partial balanced case ($|\gamma|=0.8$, $b=0.56$) in panel (h), where $F>0.82$ is maintained for $s\gtrsim 0.075\sigma$ (equivalent to 13.5 nm in typical optical microscopy).

The parallel fidelity trends in Fig.~\ref{experiment} (c), (d) [also in (g), (h)] confirm our theoretical prediction that the estimation of $r_{mn}$ is independent of source-dependent parameters ($|\gamma|$, $b$, $\varphi$). Observed deviations between MZ setup, Fig.~\ref{experiment} (c), (d), and SLM setup, Fig.~\ref{experiment} (g), (h), stem from differences in experimental noise profiles. Specifically, the MZ setup exhibits lower fidelity at small separations due to two factors: (1) Separation calibration errors, i.e., slight beam misalignments at the MZ output reduce separation measurement accuracy. (2)
Optical complexity, i.e, additional components in the MZ setup introduce background noise and image distortions.

\section{Discussion}
In conclusion, we present a unified framework for superresolution of passive two-point sources that eliminates the need to estimate source-dependent parameters, partial coherence $|\gamma|$, brightness balanceness $b$, random phase $\varphi$, and photon statistics, by leveraging the intensity/probability ratio $r_{mn}=p_m/p_n$ of even (or odd) Hermite-Gaussian modes, which depends solely on separation $s$. This approach bypasses error-prone multi-parameter estimation, reducing the problem to a single-parameter estimation and effectively decoupling $s$ from other variables.

Integrating this theory with a physics-informed convolutional neural network, we further address source-independent challenges including background luminosity noise, photon loss, centroid/orientation misalignment, via simulated training on 200,000 images ($38\times38$ pixels) with randomized parameters. Our model demonstrates exceptional robustness: (1) $>0.82$ fidelity for separations $s>0.06\sigma$ (16 times below the diffraction limit, $\sim$ 10.8 nm in optical microscopy) under full parameter uncertainty; (2) tolerance up to 20\% luminosity noise polluting over 55\% of total pixels, and up to 20\% luminosity dimming (photon loss) for over 41\% of total pixels, while maintaining high fidelity $F>0.9$; (3)
resilience against 7.5\% centroid shifts and $\pi/2$ orientation misalignment with minimal fidelity loss.

Experimental validation using Mach-Zehnder and spatial light modulator setups confirms these simulated results. For partially coherent $|\gamma|=0.8$ and unbalanced $
b=0.56$ sources, we achieve high fidelity $F>0.82$ for all $s>0.075\sigma$ (14 times below the diffraction limit, $\sim$ 13.5 nm), rivaling state-of-the-art active-source techniques.

We further show that training efficiency depends on number of training images per pixel, with even low-resolution ($10\times10$) datasets achieving $F>0.8$ for $s=0.08\sigma$ (12 times below the diffraction limit). Implemented on common commercial hardware (NVIDIA Quadro P5000 GPU), our framework trains 7 epochs in 13 minutes, highlighting its accessibility and practicality.

By bridging theoretical superresolution limits and real-world constraints, our findings enable high superresolution capabilities of uncontrollable systems, from celestial objects to live-cell structures, without prior source knowledge. The synergy of physics-driven theory and adaptive machine learning establishes a new paradigm for precision metrology, with transformative applications in astrophysics, biophotonics, and quantum sensing.

Finally, our protocol may be integrated with established PSF-reducing techniques for active sources (e.g., STED, MINFLUX, etc.), pushing superresolution limits beyond current benchmarks.  \\

\noindent{\bf Data Availability} The main data supporting the findings of this work are available within
the article and its Supplementary Information. Additional data are
available from the corresponding authors upon reasonable request.\\

\noindent{\bf Code availability} The machine learning protocol that leads to the findings of this work is described in detail in the Methods section. The python code supporting the machine learning protocol are available from the corresponding authors upon reasonable request.\\

\noindent{\bf Acknowledgements} We acknowledge partial support from NSF Grant No. PHY-2316878, the U.S. Army under Contact No. W15QKN-18-D-0040, and from Stevens Institute of Technology. \\

\noindent{\bf Author contributions} 
X.-F.Q., A.S. and B.B. conceived the research. A.S., B.B., X.-F.Q. and G.Z. developed the theory. A.S., X.-F.Q., P.K. and
G.Z. designed the experiment. A.S., P.K. and G.Z. conducted the experiments, collected and analyzed the data. X.-F.Q. and A.S. wrote the manuscript.\\

\noindent{\bf Competing interests} The authors declare no competing interests.\\


\bibliographystyle{ieeetr}


\bibliography{reference}

\section{Methods}
\subsection{Superresolution theory bypassing multi-parameter estimation}

We model a general two-point source as an arbitrary non-normalized statistical mixture of photon spatial states $\ket{h_{\pm}}$, i.e., 
\begin{equation}
\rho =p_{+}|h_+\rangle\langle h_+|+ p_{-}|h_-\rangle\langle h_-|+\sqrt{p_+p_-}\gamma|h_+\rangle\langle h_-|+\sqrt{p_+p_-}\gamma^*|h_-\rangle\langle h_+|,
\label{mixed state}
\end{equation}
where $\ket{h_{\pm}}$ represent the spatial wavefunctions of the two sources in the imaging plane, displaced by $\pm s/2$ from a Gaussian point spread function (PSF) $h(x)$ with $\la x|h_{\pm} \rangle = h_{\pm}(x) \equiv h(x\pm s/2)$. The Gaussian PSF is defined as $h^{2}(x)=\frac{1}{\sqrt{2\pi\sigma^{2}}}\exp[{-\frac{x^{2}}{2\sigma^2}}]$, where $\sigma$ represents its width. Here $s$ is the two-source separation (target parameter), $p_{\pm}$ are corresponding probabilities of the source states $|h_{\pm}\ra$ with $p_-\ge p_+$ without loss of any generality, and $\gamma=|\gamma|e^{i\varphi}$ is the coherence parameter with $|\gamma|\le 1$ due to the Cauchy–Schwarz inequality and $\varphi$ being the relative phase. We define $b=p_+/p_-\in [0,1]$ to represent the arbitrary balanceness of the two sources, where $b=1$ means balanced and $b=0$ indicates completely unbalanced. Despite the non-orthogonality of the source states $\langle h_+|h_-\rangle =\alpha \neq 0$, the density matrix $\rho$ in Eq.~\eqref{mixed state} remains fully general for describing arbitrary two-state mixture \cite{supplemental}.

According to recent superresolution approaches \cite{tsang2015quantum,tsang2016quantum,tsang2019resurgence, hradil2019quantum,hradil2021exploring, larson2018resurgence, wadood2021experimental,liang2021coherence, liang2023quantum}, one analyze the incoming generic state \eqref{mixed state} in the Hermite-Gaussian (HG) mode basis $\{\ket{\Phi_{q}}\}$,
\begin{align}
 \Phi_{q}(x)=\braket{x|\Phi_{q}}=\left(\frac{1}{2\pi\sigma^{2}}\right)^{1/4}\frac{1}{\sqrt{2^{q}q!}}H_{q}\left(\frac{x}{\sqrt{2}\sigma}\right)\exp\left[\frac{-x^{2}}{4\sigma^{2}}\right],
\end{align}
with $q=0,1,2,..,\infty$. The corresponding probability of detecting the photon in the $q$-th mode can be computed as
\begin{align}
p_{q}&=\frac{Tr[\rho|\Phi_{q}\rangle\langle\Phi_{q}|]}{Tr[\rho]}=e^{-Q}\frac{Q^q}{q!}\frac{B_q}{N(Q)}
\label{probablity}
\end{align}
where $B_q=b+1+(-1)^{q}k$, $N(Q)=b+1+ke^{-2Q}$ with $Q=s^2/16\sigma^2$ and $k=2|\gamma|\sqrt{b}\cos{\varphi}$. 

Apparently, the detection of probability $p_q$ depends on four realistically unknown parameters: separation $s$ via $Q$, partial coherence $|\gamma|$, balanceness $b$, and relative phase $\varphi$, i.e., $p_q\equiv p_q(s,|\gamma|,b,\varphi)$. Estimating $s$ from $p_q$ thus becomes a multi-parameter estimation problem, which severely degrades the Fisher information \cite{arvidsson2020quantum, helstrom1969quantum, liu2020quantum} and introduces significant uncertainty. Even in simplified cases, such as estimating only $s$ and $|\gamma|$, this parameter coupling persists, as shown in Ref.~\cite{larson2018resurgence}, underscoring the inherent limitations of conventional approaches for realistic passive-source superresolution.

In the idealized case of incoherent ($|\gamma|=0$) and balanced ($b=1$) sources, the detection probability $p_q$ reduces to a single-parameter dependence on $s$. This simplification preserves non-zero Fisher information \cite{tsang2016quantum}, enabling reliable separation estimation, a requirement satisfied only under these conditions. Consequently, prior passive-source superresolution methods remained confined to this restrictive regime, where parameter coupling vanishes.

To circumvent the multi-parameter estimation challenge, we introduce a parameter-decoupling protocol based on the ratio of even (or odd) HG mode probabilities. Critically, this ratio depends solely on the separation $s$,
\begin{equation}
r_{mn}(s)=\frac{p_{m}}{p_{n}}=\frac{\eta_{m} (s)}{\eta_{n}(s)},
\label{ratio}
\end{equation}
where $m,n=0,2,4,...$, are arbitrary even integer numbers and $\eta_{q}(s)=e^{-Q}\frac{Q^q}{q!}$ depends only on the separation $s$, decoupling it from $|\gamma|$, $b$, and $\varphi$. This reduces the problem to single-parameter estimation, enabling direct computation of $s$:
\begin{align}
\label{s-ratio}
s&=4\sigma\left[\frac{m!r_{mn}}{n!}\right]^{1/2(m-n)}.
\end{align}
The flexibility in choosing $m$ and $n$ accommodates diverse experimental configurations. For instance, selecting $m=2$ and $n=0$ yields $s=4\sigma(2r_{20})^{1/4}$.

Additionally, the probability ratio $r_{mn}$ is also independent of the photon statistics or photon number of the source. This justifies that the Fisher information, when decomposed into HG modes \cite{tsang2016quantum}, needs only to consider the variation of separation $s$, and can be computed as
\beqa
I^{(\mathrm{HG})}(s)&=& \frac{Q}{4\sigma^2} \frac{1+b}{N(Q)} \left((1+A(Q))^2 + \frac{1}{Q} \right) + \frac{Q}{4\sigma^2} \frac{k e^{-2Q}}{N(Q)} \left( (1-A(Q))^2  - \frac{1}{Q}\right) 
\label{FisherInfo}
\eeqa
where $A(Q)=[2ke^{-2Q}-N(Q)]/N(Q)$ \cite{supplemental}. As shown in Fig.~\ref{roadmap} (d), $I^{HG}(s)$ vanishes only under the extreme condition of complete destructive interference, i.e., complete spatial overlap  $s=0$, full coherence $|\gamma|=1$, balanced $b=1$, and fixed relative phase $\varphi=\pi$. For all other cases, which is practically always true in real-world situations, one consistently has $I^{HG}(s)>0$. Thus, this Fisher information can be considered as practically non-vanishing even when $s\rightarrow 0$, ensuring reliable estimation even at sub-diffraction scales.

It is important to note that the above formalism extends to classical partially coherent two-point sources by replacing quantum probabilities with classical intensities. The ratio $r_{mn}$ and Fisher information $I^{(HG)}$ retain identical forms, ensuring applicability to both quantum and classical regimes \cite{supplemental}.


\subsection{Physics-informed machine learning}

In general, directly measuring the intensity/probability ratio $r_{mn}(s)$ is challenging and, in some cases, impractical or impossible. Furthermore, direct measurement of decomposed mode ratios cannot address additional practical source-independent challenges including background noise, detection inefficiencies (e.g., photon loss), two-source center point and orientation misalignment, and other related factors.

To tackle these two layers of challenges, we employ a convolutional neural network (CNN) as a versatile tool to achieve two goals: (i) obtaining the ratio $r_{mn}$ for pairs of spatial modes
by training images based on the analytical relation \eqref{ratio} with arbitrary source-dependent imperfections $|\gamma|$, $b$, $\varphi$, and (ii) mitigating the adverse effects of practical source-independent imperfections through training prescriptions based on simulated realistic physical scenarios. Fig.~\ref{roadmap} (b) provides a schematic illustration of our CNN procedure.

To achieve goal (i), we take the first six modes to construct probability ratios \eqref{ratio} in the training process. This is due to the fact that two closely separated point sources are mostly concentrated in lower order HG spatial modes. It is realized through creating six neurons (employing ReLU activation function) in the last dense layer that connects to the flattened one-dimensional vector generated from a series of 2D-convolutional layers. These six neurons utilizes the sigmoid function to produce output values within the range of 0 to 1 to create mutual contribution ratios among the six modes. We remark that the using of six modes is already achieving high fidelity of estimating the separation, but more modes will definitely further improve the output quality. The training images are then generated by setting parameters $|\gamma|$, $b$, and $\varphi$ as random numbers to account for arbitrary partial coherence, balanceness, and relative phase. The separation $s$ is also generated randomly, particularly focusing the range below conventional resolution limit.

To accomplish goal (ii), we design special training procedures. To account for background noise, we generate training images with random extra  brightness (up to 20\% of the signal) for a number of random pixels. Similarly, to include possible detection inefficiency or photon loss, we generate training images with random extra dimness (up to 20\% less of the signal) for a number of random pixels. The center-point misalignment is addressed by introducing a random variation of the centroid function $h(x)$ for all training images. The two-source orientation misalignment is accounted by using training images with random orientations. 

Accounting all factors described above in realizing goals (i) and (ii), we produce a dataset of 200,000 training images with $38\times38$ pixels. The CNN model gradually increases the number of filters through its convolutional layers and concludes with a multi-class classification output layer, making it suitable for identifying complex patterns or classes within images while maintaining sensitivity to intricate image features. We carry out the downsizing using Average-pooling layers instead of Max-pooling (which is commonly used), in order to keep the profile information while downsizing the image by selecting maximum pixel values in the pooling region. To ensure high accuracy we employed seven epochs during training. Fig.~\ref{roadmap} (c) illustrates examples of training images.

These training steps are primarily facilitated by the theoretical parameter-decoupled superresolution results \eqref{probablity}-\eqref{s-ratio}, as well as by realistic physical contexts. Therefore, we refer to our specialized machine learning procedure as a Physics-Informed CNN. The training and testing processes are conducted using a regular commercial workstation equipped with an Intel Xeon Platinum 8160 CPU operating at a base frequency of 2.10 GHz. The CPU comprises 48 cores and 96 logical processors. The system also utilizes an NVIDIA Quadro P5000 GPU with 16 GB of dedicated memory and 144 GB of total GPU memory. The training of seven epochs takes about 13 minutes.

In the results section, two sets of testing data are reported to verify the performance of our physics-informed CNN model, i.e., computer-generated testing images and lab-generated testing images.

\subsection{Experimental scenarios}

To demonstrate the broad applicability of our physics-informed CNN super-resolution model, we employ two distinct experimental methodologies to generate a diverse set of testing two-source images under varying conditions, i.e., by a modified Mach-Zehnder interferometer, and a spatial light modulator. These two scenarios introduce different types of realistic imperfections and accompanied systematic noises, providing a more rigorous and challenging test for the CNN model.

{\bf Experimental scenario (1):} In this experimental configuration, we employ a Mach-Zehnder interferometer featuring a mirror mounted on a translation stage within one interferometric arm to create two parallel output beams with controllable separation $s$ and balanceness $b$. As shown in Fig.~\ref{experiment} (a), a polarized 637 nm CW beam is sent through a half wave plate (HWP) for polarization rotation before splits into two by a polarizing beam splitter (PBS). The combination of HWP and PBS controls the two-path relative intensities/probabilities, or the balanceness $b$ of the two sources at the output. Mirror M1, mounted on a translation stage T in path 1, is dynamically adjusted to generate varying separation $s$ of its beam from the one exits path 2 at the output. The collinear propagation of the two beams after the regular 50/50 beam splitter is realized by the adjustments of mirrors M2 and M3. The source image is then captured by the CCD camera for testing the CNN model. The use of Shutter $S$ helps to block the beam in path 2 and calibrate the actual distance produced by the translation stage T between the two output sources.

{\bf Experimental scenario (2):} In this scenario, we employ a spatial light modulator (SLM) to construct a two-source signal with varying partial coherence $|\gamma|$, balanceness $b$, relative phase $\varphi$, and separation $s$. As shown in Fig.~\ref{experiment} (e), the two source signal is created by the Hamamatsu LCOS SLM through Computer Generated Hologram (CGH). The reflection from the SLM is then focused and collimated using a 4F lens system for imaging. The desired order of the diffraction pattern that contains the encoded beam is selected with a pinhole before collimation. Partial coherence is equivalently realized by a relative phase modulation, see an analysis of the equivalence in Ref.~\cite{hradil2019quantum}. The distance between the centers of the two beams is also encoded to change with step size around $\sigma/50$. Then the two-source signal is recorded for testing by a CCD camera.

In both experimental implementations, several challenges arise for the quality of the captured images. First, background noise inherent to the imaging system and environmental conditions introduces uncertainty in the recorded intensities, potentially obscuring subtle features required for precise analysis. Second, while the Gaussian beams are theoretically expected to exhibit a perfect Gaussian profile, practical distortions in the beam shape, due to imperfections in optical components or alignment errors, result in deviations from the ideal form. These distortions can impact the accuracy of the separation estimation. Third, the alignment of the two beams presents another significant challenge, as deviations from parallel propagation lead to angular misalignment. It not only complicates the interpretation of the separation but also increases the sensitivity to experimental errors during collimation and imaging. All these realistic imperfection factors provide an ideal testbed in validating the robustness and reliability of the proposed physics-informed CNN superresolution methodology.

\end{document}